\documentclass[prb,showpacs,amsmath,amssymb,twocolumn,
superscriptaddress]{revtex4}
\usepackage{graphicx}
\usepackage{dcolumn}
\usepackage{bm}
\usepackage{mathrsfs}
\usepackage{subfigure}
\usepackage{natbib}
\usepackage{amssymb}
\usepackage{multirow}
\usepackage{float}
\usepackage{color}
\usepackage{bbm}
\usepackage{threeparttable}
\begin{document}

\title{Interplay of Fermi velocities and healing lengths in two-band superconductors}

\author{Yajiang Chen}
\affiliation{Department of Physics, Zhejiang Sci-Tech University, 310018 Zhejiang, China}

\author{Haiping Zhu}
\affiliation{Department of Physics, Lishui University, 323000 Zhejiang, China}

\author{A. A. Shanenko}
\affiliation{Departamento de Fisica, Universidade Federal de Pernambuco, Cidade Universitaria, 50670-901 Recife-PE, Brazil}

\date{\today}
\begin{abstract}
By numerically solving the Bogoliubov-de Gennes equations for the single vortex state in a two-band superconductor, we demonstrate that the disparity between the healing lengths of two contributing condensates is strongly affected by the band Fermi velocities, even in the presence of the magnetic field and far beyond the regime of nearly zero Josephson-like coupling between bands. Changing the ratio of the band Fermi velocities alters the temperature dependence of the condensate lengths and significantly shifts parameters of the ``length-scales locking" regime at which the two characteristic lengths approach one another.
\end{abstract}

\maketitle

\section{Introduction}
\label{sec1}

Characteristic length scales associated with different contributing condensates constitute one of the cornerstone features of multiband superconductors. Multiple condensates in one system interfere, which results in unconventional coherent phenomena~\cite{Milorad2015}. Effects of such interference are most pronounced when the spatial lengths of the contributing condensates are notably different. Various definitions of such lengths are in use, including those related to the gap function slope in the vortex core~\cite{Caroli1964, Gygi1991}, the maximum density of the supercurrent~\cite{Sonier2004}, the radius of a cylinder containing energy equal to the condensation energy~\cite{Gennes1966, Tinkham1996}, or the healing length along which the condensate reaches $60$-$80\%$~(there are different choices) of its bulk value~\cite{Komendova2011, Komendova2012, Saraiva2017}. All such definitions produce similar results (except of the slope definition that fails at nearly zero temperatures due to the Kramer-Pesch collapse~\cite{Kramer1974,Chen2014}) and either of them can be employed to characterize the condensate spatial scales.

Since 1970s it is well known~\cite{Geilikman1967} that the spatial lengths of different band condensates in multiband materials are the same in the Ginzburg-Landau (GL) domain, see also Refs.~\onlinecite{Koshelev2004, Koshelev2005, Geyer2010, Kogan2011, Shanenko2011, Vagov2012}. However, using the perturbative expansion of the microscopic equations in the small deviation from the critical temperature $T_c$ to one order beyond the GL theory (extended GL), one finds that the band-dependent condensate lengths can be different~\cite{Shanenko2011,Vagov2012,Saraiva2017}. This conclusion was confirmed by numerically solving the two-band Bogoliubov-de Gennes (BdG) equations~\cite{Komendova2012}. Moreover, it was also demonstrated~\cite{Komendova2012} that the condensate characteristic length associated with a weaker band notably increases when approaching the critical temperature of this band taken as a separate superconductor (the hidden critical point). As the length of the stronger band condensate remains unaffected in this case, one can get an increased difference between the two lengths governed by the hidden criticality. On the other hand, the condensate lengths for sufficiently strong interband couplings tend to be nearly the same, as demonstrated in Refs.~\onlinecite{Saraiva2017} and \onlinecite{Ichioka2017}. This can possibly explain the recent scanning tunneling microscopy measurements~\cite{Fente2016} and can be referred to as the ``length-scales locking"~\cite{Ichioka2017}.

Though the disparity between the condensate lengths is more pronounced for weaker interband couplings, the ``length-scales locking" regime can be shifted toward larger values of the interband couplings. Indeed, it was recently shown within the extended GL formalism~\cite{Saraiva2017} that the difference between the condensate healing lengths in a two-band superconductor is very sensitive to the ratio of the band Fermi velocities $v_{Fi}\;(i=1,2)$ which varies within a wide range, see, for example, Table~\ref{tab1} illustrating some experimental results.
\begin{table}[!h]
\renewcommand\arraystretch{1.5}
\caption{The band Fermi velocities $v_{F1}$ and $v_{F2}$ of two-band superconductors in the units of $10^5$ m/s. The indices $1$ and $2$ correspond to the stronger and weaker bands, respectively.}\label{tab1}
\setlength{\tabcolsep}{1.2mm}{
\begin{threeparttable}
\begin{tabular}{ccccc}
  \hline
Material & $v_{F1}$ & $v_{F2}$ & $v_{F2}/v_{F1}$ & Ref.\\ \hline\hline
2H-NbS$_2$ & 3.1 & 0.155 & 0.05 &Ref.~\onlinecite{Tissen2013}\\
Ba$_{0.85}$K$_{0.15}$Fe$_2$As$_{2}$  & - & - & 0.10\tnote{*} & Ref.~\onlinecite{Tarantini2011}\\
Ba$_{0.6}$K$_{0.4}$Fe$_2$As$_{2}$  & - & - & 0.95 & Ref.~\onlinecite{Tarantini2011}\\
MgB$_2$ & 4.4 & 8.2 & 1.86 & Ref.~\onlinecite{Knight2008}\\
2H-NbSe$_2$ & 0.55 & 10 & 18.2 & Ref.~\onlinecite{Suderow2005}\\
  \hline
  \hline
\end{tabular}
    \begin{tablenotes}
        \footnotesize
        \item[*] Extracted from the upper critical field $H_{c2}$, with $H$ parallel to the $c$ axis.
      \end{tablenotes}
    \end{threeparttable}
}
\end{table}
Furthermore, this ratio can be altered by doping in superconductors~\cite{Lee2006}, changing the topology of the Fermi surface~\cite{Cappelluti2007}, engineering the interface of the system~\cite{Peng2014}, applying the pressure~\cite{Tissen2013, Suderow2005}, and changing the characteristic size of nanoscale superconductors via the quantum-size effects~\cite{Blatt1963, Shanenko2006a, Guan2010, Chen2010,Shanenko2008,Chen2012a}. As the impact of the Fermi velocities on the condensate lengths was investigated by means of the extended GL formalism and in the absence of the magnetic effects, it is of importance to compliment the conclusions of Ref.~\onlinecite{Saraiva2017} by investigating temperatures far below $T_c$ and including the magnetic field.

In this work we explore the interplay between the band Fermi velocities and the condensate healing lengths by numerically solving the two-band BdG equations for a single vortex solution in the entire range of the temperatures below $T_c$. As the local magnetic field is not neglected, the BdG equations are supplemented by Ampere's law introducing an additional magnetic coupling between the contributing condensates.
The special attention is also given to the effect of the hidden criticality at which the disparity between the healing lengths is most pronounced.

The paper is organized as follows. In Sec.~\ref{form}, we outline the formalism of the BdG equations for a single-vortex state in a two-band condensate. The numerical results and related discussions are given in Sec.~\ref{res} including three subsections. The first subsection presents results for the zero temperature $T = 0$ and zero external field $H =0$. Here one can find the healing lengths $\xi_1$ and $\xi_2$ as functions of the Fermi velocities ratio $v_{F2}/v_{F1}$ and the interband coupling $g_{12}$. For illustration, we also show how the healing lengths are extracted from the spatially dependent gap functions. The results for $T \neq 0$ and $H = 0$ are discussed in the second subsection. Here we investigate the healing lengths as functions of $T$ for different parameters $v_{F2}/v_{F1}$ and $g_{12}$. In the third subsection we discuss the results for $T \neq 0$ and $H \neq 0$. Conclusions are given in Sec.~\ref{conclusions}.

\section{Formalism}
\label{form}

To investigate how the spatial scales of the partial band condensates in a two-band superconductor are sensitive to the band Fermi velocities, a single vortex solution of the two-band BdG equations is considered in a cylinder with the vortex line parallel to the $z$ axis of this cylinder. We utilize the standard microscopic model of a two-band superconductor~\cite{Suhl1959,Moskal1959} with the conventional $s$-wave pairing in both bands, controlled by the symmetric coupling matrix $g_{ii'}$~($i,i'=1,2$). The intraband couplings $g_{11}$ and $g_{22}$ are chosen so that the critical temperature of band $1$, taken as a separate superconductor, is larger than the critical temperature in the decoupled band $2$, i.e., we have stronger band $1$ and weaker band $2$. The two condensates are coupled through the Josephson-like transfer of Cooper pairs controlled by $g_{12}$. The parabolic single-particle energy dispersion is assumed for charge carriers in both bands. For our calculation we choose quasi-2D bands, as multiband materials often exhibit quasi-2D Fermi surfaces, see e.g. Ref.~\onlinecite{Paglione2010}. An external magnetic field is applied along the $z$ axis of the cylinder while the dependence of the quasi-2D band dispersions on the $z$ projection of the single-particle momentum is minor and neglected in our calculations. The superconductor is in the clean limit.

The corresponding BdG equations read~\cite{Komendova2012,Araujo2009}
\begin{equation}\label{bdg}
\left[
\begin{array}{cc}
\hat{H}_{ei} & \Delta_i({\bf r}) \\
\Delta_i^*({\bf r}) & -\hat{H}^*_{e,i}
\end{array}
\right] \left[
\begin{array}{c}
u_{i\nu}({\bf r}) \\
v_{i\nu}({\bf r})
\end{array}
\right] = E_{i\nu}
\left[
\begin{array}{c}
u_{i\nu}({\bf r}) \\
v_{i\nu}({\bf r})
\end{array}
\right],
\end{equation}
where $u_{i\nu}(\bf r)$ and $v_{i\nu}({\bf r})$ are the electron-like and hole-like wave functions associated with band $i$ ($\nu$ is the set of the relevant quantum numbers); $E_{i\nu}$ and $\Delta_i({\bf r})$ are the corresponding quasiparticle energy and the spatial pair potential (gap function); and the single-particle Hamiltonian for the charge carriers in band $i$ is given by
\begin{align}
\hat{H}_{e,i}({\bf r})=-\frac{\hbar^2{\bf D}^2}{2m_i}  - \mu_i,
\label{T}
\end{align}
with $m_i$ the electron band mass, $\mu_i=m_iv_{Fi}^2/2$ the chemical potential measured from the lower edge of the corresponding band, ${\bf D}= \boldsymbol{\nabla} - \mathbbm{i}\frac{e}{\hbar\mathbbm{c}} {\bf A}$, and ${\bf A}({\bf r})$ the vector potential.

As the problem is solved in a self-consistent manner, the band gap functions and the vector potential depend on the solutions of Eqs.~(\ref{bdg}) as
\begin{align}
 \Delta_i({\bf r}) = &\sum_{i'\nu} g_{ii'}\, u_{i'\nu}({\bf r})
v^*_{i'\nu}({\bf r})\big[1-2f(E_{i'\nu})\big]\label{op}
\end{align}
and
\begin{align}
\boldsymbol{\nabla}\times \boldsymbol{\nabla}\times {\bf A}({\bf r})= \frac{4\pi}{\mathbbm{c}} {\bf j}({\bf r}),
\label{ampere}
\end{align}
where $f(E_{i'\nu})$ is the Fermi-Dirac distribution and the supercurrent density is given by
\begin{align}
{\bf j}({\bf r}) =
&\sum_{i'\nu} \frac{e\hbar}{2m_{i'}\mathbbm{i}}\Big\{ f(E_{i'\nu})\, u^*_{i'\nu}({\bf r}) {\bf D} u_{i'\nu}({\bf r})\notag \\
&+\big[1-f(E_{i'\nu})\big]\, v_{i'\nu}({\bf r}){\bf D} v^*_{i'\nu}({\bf r}) -{\rm h.c.}\Big\}.\label{current}
\end{align}
The summation in Eqs.~(\ref{op}) and (\ref{current}) goes over the quasiparticle states with positive energies. In addition, Eq.~(\ref{op}) includes only the states for which the averaged single-electron energy taken at zero field~\cite{Shanenko2008} $\langle \hat{H}_{e,i} \rangle|_{{\bf A} =0}$ falls into the range $[-\hbar \omega_D,~\hbar\omega_D]$, with $\omega_D$ the Debye frequency assumed the same for both contributing bands. Similar results (with deviations of about $1$-$2\%$) can be obtained when selecting $E_{i \nu} < \hbar\omega_D$ in Eq.~(\ref{op}), see e.g. Ref.~\onlinecite{Gygi1991}.

The Josephson-like coupling between the two contributing bands is not explicitly present in the Bogoliubov-de Gennes Eqs. (\ref{bdg}),  appearing in the self-consistency gap equation Eq. (\ref{op}). The magnetic coupling between the condensates manifests itself through the presence in Eqs.~(\ref{bdg}) of the vector potential that is related to the both contributing condensates by means of Ampere's law Eq. (\ref{ampere}). We remark that to go beyond the adopted model, the pairing of electrons from different bands should be taken into consideration, i.e. in addition to the transfer of the Cooper pairs from one band to another, one accounts for an extra coupling through the interband Cooper pairs, including one electron from band $1$ and another from band $2$, see e.g. Ref.~\onlinecite{Shanenko2015}. In this case the coupling between bands appears in the Bogoliubov-de Gennes equations~\cite{Shanenko2015,Vargas2020}. However, in most cases the interband pairing is suppressed due to incommensurability of the Fermi momenta in different bands and can be neglected.

Considering a single vortex oriented along the $z$ direction, we follow the previous studies of a single vortex solution within the single-band
\cite{Gygi1991, Bardeen1969, Hayashi1998} and two-band BdG equations~\cite{Komendova2012,Araujo2009}. Due to the cylindrical geometry, we can write
\begin{eqnarray}
\Delta_i({\bf r}) &=& \Delta_i(\rho)e^{-i\theta}, \label{op2}
\end{eqnarray}
and
\begin{eqnarray}
u_{i\nu }({\bf r}) &=& \frac{1}{\sqrt{2\pi
L}}u_{ijm}(\rho)e^{i(m-\frac{1}{2})\theta}e^{\mathbbm{i}k_zz}, \nonumber \\
v_{i\nu}({\bf r}) &=& \frac{1}{\sqrt{2\pi
L}}v_{ijm}(\rho)e^{i(m+\frac{1}{2})\theta}e^{\mathbbm{i}k_zz},\label{uv2}
\end{eqnarray}
where $\rho,\theta$ and $z$ are the cylindrical coordinates, $L$ is the unit cell of the periodic boundary conditions in the $z$-direction, and $\nu=\{j, m, k_z\}$ with $j$ the radial quantum number, $m$ the azimuthal quantum number being half an odd integer, and $k_z$ the wavenumber in the $z$-direction. As mentioned above, the dependences of the quasi-2D band dispersions on $k_z$ are neglected and so, the wave functions $u_{i\nu }({\bf r})$ and $v_{i\nu }({\bf r})$ are not dependent on $k_z$ either.

For the chosen gauge and symmetry, ${\bf A}({\bf r}) = A_{\theta}(\rho){\bf e}_{\theta}$, with ${\bf e}_{\theta}$ the azimuthal unit vector. The two boundary conditions for $A_{\theta}(\rho)$ are set as: (1) the magnetic field approaches the external one $H{\bf e}_{z}$ far away from the cylinder; (2) the magnetic field is finite at the origin of the coordinates (the vortex center). The latter assumes $A_{\theta}(0) = 0$ to avoid the divergence of the field. In addition, the transverse quantum confinement requires the boundary conditions $u_{ijm}(\rho=R)=0$ and $v_{ijm}(\rho =R) = 0$, where $R$ is the radius of the cylinder.

To represent the BdG equations in the matrix form, we expand the radial parts of the particle-like and hole-like wave functions $u_{i jm}(\rho)$ and $v_{i jm}(\rho)$ in terms of the normalized Bessel functions of the first kind
\begin{equation}
\phi_{im}^{(\pm)}(\rho)= \frac{\sqrt{2}}{R\mathcal{J}_{(m+1)\pm\frac{1}{2}}(\alpha_{i,m\pm\frac{1}{2}})}
\mathcal{J}_{ m\pm\frac{1}{2}} \big(\alpha_{i,m\pm\frac{1}{2}}\frac{\rho}{R}\big),
\label{phi}
\end{equation}
where superscripts ``-" and ``+" are for $u$ and $v$ functions, respectively, and $\alpha_{i,\eta}$ is the $i$th zero of the corresponding Bessel function, i.e., $J_{\eta}(\alpha_{i,\eta})=0$. The expansion writes
\begin{align}
u_{ijm}(\rho) &=\sum\limits_{i=1}^N c_{ijj'm}\phi^{(-)}_{j'm}(\rho),\notag\\
v_{ijm}(\rho) &=\sum\limits_{i=1}^N d_{ijj'm}\phi^{(+)}_{j'm}(\rho) ,
\label{exp}
\end{align}
where $N$ should be chosen sufficiently large to capture the essential features of the vortex solution. As a result, the BdG equations are reduced to the matrix ($2N\times 2N$) equation with the eigenvectors given by $\{c_{ijj'm}\}$~(upper half of the column) and $\{d_{ijj'm}\}$ (lower half). Then, the numerical solution of the problem is calculated in the self consistent manner. First, we choose some initial gap functions $\Delta_i(\rho)$ and vector potential $A_\theta(\rho)$ and find the corresponding eigenvalues and eigenstates of the matrix BdG equations. Second, we use the obtained sets $\{c_{ijj'm}\}$ and $\{d_{ijj'm}\}$ and the related quasiparticle energies $E_{ijm}$ to calculate the new position dependent gaps and vector potential by means of the equations Eqs.~(\ref{op})-(\ref{current}), (\ref{uv2}), and (\ref{exp}). Third, the BdG equations are solved again with the calculated gap functions and vector potential. The procedure is repeated until the convergence.

Below the effective band-dependent electron masses $m_i$ are set to the free electron mass $m_e$, for simplicity. The intraband couplings are chosen such that $g_{11}N_1 = 0.3$ and $g_{22}N_2 = 0.24$, where $N_i$ is the normal density of states (DOS) per spin projection of band $i$. In the case of interest $N_1=N_2=(m_e/2\pi\hbar^2 L) \sum_{k_z}\theta(k_{\rm max} - |k_z|)$, with $\theta(k_{\rm max} - |k_z|)$ the step function and $k_{\rm max}$ the maximal wavenumber in the $z$ direction. One can estimate $k_{\rm max} = \pi/a_z$, where $a_z$ is the corresponding lattice constant. Then, using the periodic boundary conditions for the motion in the $z$ direction, one gets $(1/L) \sum_{k_z}\theta(k_{\rm max} - k_z) \sim 1/a_z$. Keeping in mind typical values for the lattice constant, one concludes that $1/a_z$ is of the order of $1$-$3 {\rm nm}^{-1}$. For our calculations we choose $N_1=N_2= \tilde{N} m_e/ 2\pi\hbar^2$, with $\tilde{N}=1 {\rm nm}^{-1}$~(similarly to Ref.~\onlinecite{Gygi1991}). Notice that this choice and also the use of $m_i=m_e$ do not influence our conclusions because any changes in $N_i$ result simply in the renormalization of the intraband couplings $g_{11}$ and $g_{22}$, as we keep the same dimensionless couplings $g_{11}N_1$ and $g_{22}N_2$. Notice that the chosen values for the intraband couplings are in the typical range for multiband materials, see Ref.~\onlinecite{Vagov2016} and references therein. The interband coupling $g_{12}$ is varied in our study, in order to investigate effects of the interaction between the two contributing condensates.

To have different Fermi velocities $v_{F1}$ and $v_{F2}$, we choose different $\mu_1$ and $\mu_2$. For the stronger band we adopt $\mu_1 = 30$ meV, based on conservative estimates of the Fermi energy in emergent multiband  superconductors, see e.g. Ref.~\onlinecite{Lubashevsky2012}. The chemical potential relative to the lower edge of the weaker band $\mu_2$ is varied in our calculations so that the ratio of the band Fermi velocities $v_{F2}/v_{F1}$ is altered by this variation.

To avoid effects of quantum confinement, the radius $R$ of the cylinder should be chosen sufficiently large. When taking the Debye frequency as $\hbar\omega_D = 15$ meV (in the range of conventional values, see e.g. Ref.~\onlinecite{Fetter2003}), one finds that for the zero temperature the healing lengths $\xi_1$ and $\xi_2$ are not sensitive to the cylinder radius for $R \gtrsim 100\,{\rm nm}$. For example, the calculations yield $\xi_1 = 19.2$ nm and $\xi_2=29.3$ nm when employing $g_{12} = 0.05g_{11}$ and $v_{F2}/v_{F1}=1$ for $T,H=0$. Then, choosing $R=300\, {\rm nm}$, we safely have $R > \xi_{1,2}$ up to the temperatures $T\approx 0.99T_c$. Since the healing lengths increase with the temperature approximately as~\cite{Fetter2003} $\propto \tau^{-1/2}$, with $\tau=1-T/T_c$,
they approach $R$ at $T\approx 0.99T_c$. Only in this case $\xi_1$ and $\xi_2$ are affected by the geometry of the sample.

We also note that the presence of the boundary conditions $u_{ijm}(\rho=R)=0$ and $v_{ijm}(\rho =R) = 0$ introduces an additional condensate length near the boundary. Indeed, here $\Delta_i(\rho)$ exhibits a series of the Friedel-like oscillations~\cite{Troy1995,Martin1997} with the period of a half of the band-dependent Fermi wavelength $\lambda_{Fi}/2$. For the chosen parameters we have $\lambda_{F1}/2=1.1\, {\rm nm}$ and $\lambda_{F2} \sim \lambda_{F1}$. One sees that  $\lambda_{F1,2}/2$ is much smaller than $\xi_{1,2}$  and the presence of the Friedel-like oscillations can in no way distort our results.

\section{Results and Discussion}
\label{res}

In this section we discuss the results of numerically solving the BdG equations for a single vortex in the two-band superconducting condensate within the model outlined in the previous section.

Before the discussion, we need to stress that the case of the zero external field $H=0$ does not assume the absence of any magnetic effects. The local field $B$ is always nonzero in the vortex core and the magnetic coupling between the two band condensates is present even for $H=0$ due to Ampere's law given by Eq.~(\ref{ampere}). [Obviously, its impact on the healing lengths can be neglected only in deep type II.] Then, the question arises which boundary conditions for the magnetic field far beyond the vortex core we should use, to obtain relevant information about the condensate healing lengths in the mixed state. We recall that for a single-vortex solution in bulk we have $B \to 0$ at infinity, see e.g. Ref.~\onlinecite{Vagov2016}. Furthermore, near the lower critical field $H_{c1}$, an Abrikosov lattice exhibits a significant distance separating neighbouring vortices so that the single-vortex state is a good approximation for such a dilute lattice. In this case the local field $B$ is indeed exponentially small between vortices, being far smaller than the external field. Clearly, to describe this case, the boundary condition $B \to H=0$ should be applied far beyond the vortex core in our calculations.

Near the upper critical field $H_{c2}$ the local field $B$ approaches the external magnetic field between vortices in the vortex matter. We can model this situation by invoking the boundary condition $B \to H\not= 0$ far beyond the vortex core. However, it is necessary to keep in mind that the healing lengths for the single-vortex state can deviate from the corresponding lengths in a dense Abrikosov lattice appearing close to $H_{c2}$. Below we investigate both $H=0$ and $H\not= 0$. We expect that the former case gives the healing lengths in the two-band superconductor near the lower critical field while the latter case is more suitable to consider the mixed state near the upper critical field.

\subsection{Zero $T$ and $H=0$}
\label{sub1}

\begin{figure}[t]
\centering
\includegraphics[width=1\linewidth]{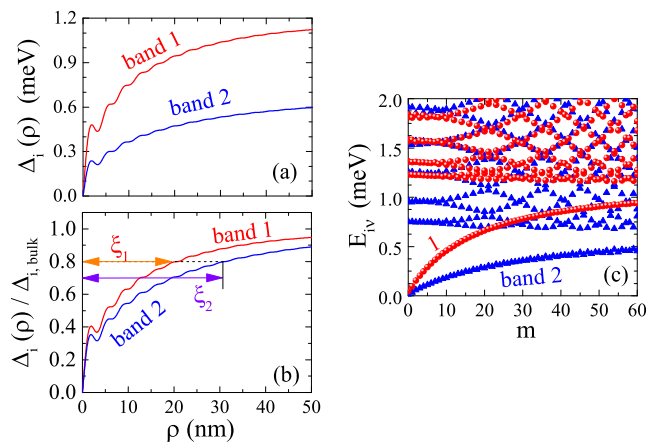}
\caption{(Color online) The single vortex solution for $v_{F2}/v_{F1}=1$ and
$g_{12} = 0.05g_{11}$ at $T = 0$ and $H = 0$: (a) $\Delta_i(\rho)$ versus $\rho$ for two bands $i=1,2$, (b) the normalized gap functions $\Delta_{i}(\rho)/\Delta_{i,{\rm bulk}}$ as functions of $\rho$ , and (c) the quasiparticle energies $E_{i\nu}=E_{ijm}$ versus the azimuthal quantum number $m$.}
\label{fig1}
\end{figure}

Our starting point is the case $T,H=0$. First we discuss how the healing lengths are extracted from the numerical results. Figure~\ref{fig1}(a) demonstrates the position dependent gap functions $\Delta_1(\rho)$ and $\Delta_2(\rho)$ calculated for $g_{12} = 0.05g_{11}$ and $v_{F2}/v_{F1}=1$. Figure~\ref{fig1}(b) shows the same gap functions but normalized by their bulk values $\Delta_{i,{\rm bulk}}$. This panel of Fig.~\ref{fig1} also illustrates the procedure of extracting the related healing lengths. For convenience of the reader, the corresponding quasiparticle spectrum $E_{i\nu}=E_{ijm}$ is shown as a function of the azimuthal quantum number $m$ in Fig.~\ref{fig1}(c)~[the data for band $1$ are given by circles while band $2$ is represented by the triangles].

In Figs.~\ref{fig1}(a) and (b), one can see fast spatial oscillations with the period $\lambda_{Fi}/2$ inside the vortex core, similarly to the single-band case~\cite{Hayashi1998}. As $v_{F2}/v_{F1} = 1$, the period of such oscillations is the same for both contributing bands. Their appearance in low-temperature clean superconductors is related to the Kramer-Pesch collapse~\cite{Kramer1974,Chen2014} of the vortex core. In this case each condensate exhibits two spatial scales: the short (anomalous) one is governed by $\lambda_{Fi}/2$ and another is related to the condensate healing lengths $\xi_i$. At zero temperature one cannot extract $\xi_i$ from the gap function slope affected by the anomalous spatial scale. However, the short scale oscillations exist only at nearly zero temperatures and are washed out above $0.1T_c$. For larger temperatures all the definitions of the condensate characteristic length, mentioned in the Introduction, produce similar results. In our work, to extract the band-dependent healing lengths, we adopt the criterion $\Delta_i(\rho=\xi_i) = 0.8 \Delta_{i,{\rm bulk}}$, see Fig.~\ref{fig1}(b) and Ref.~\onlinecite{Komendova2012}.

In Fig.~\ref{fig1}(c) one sees the bound (in-gap) quasiparticle states for each band that are responsible for the deviations of $\Delta_i(\rho)$ from its bulk value and, thus, control the condensate healing lengths $\xi_i$.

\begin{figure}[t]
\centering
\includegraphics[width=1\linewidth]{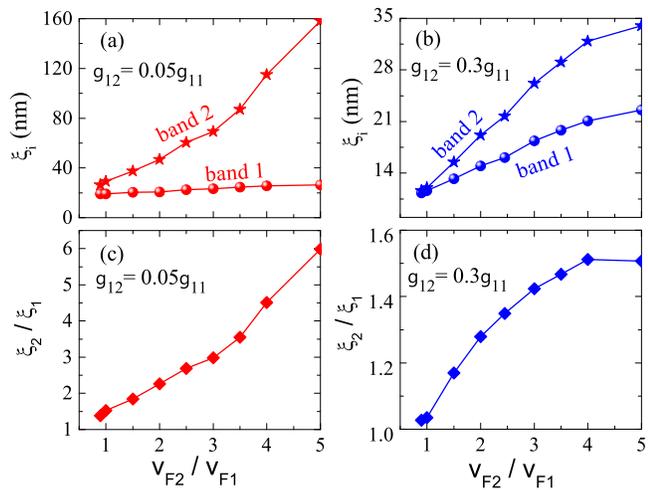}
\caption{(Color online)  The healing lengths $\xi_1$ and $\xi_2$ as functions of $v_{F2}/v_{F1}$
at $T,H = 0$, as calculated for $g_{12}=0.05g_{11}$~(a) and $0.3g_{11}$~(b). The corresponding ratio $\xi_{2}/\xi_1$ versus $v_{F2}/v_{F1}$ for the weaker (c) and larger (d) interband couplings.}
\label{fig2}
\end{figure}

In Figs.~\ref{fig2}(a) and (b) the dependence of $\xi_1$ and $\xi_2$ on the Fermi velocities ratio $v_{F2}/v_{F1}$ is shown for two values of the interband coupling $g_{12}=0.05g_{11}$ and $g_{12}=0.3g_{11}$. One can see that the both healing lengths increase with $v_{F2}/v_{F1}$. However, $\xi_2$ is much more sensitive to the value of this ratio. In particular, when $v_{F2}/v_{F1}$ goes from $1$ to $5$ in Fig.~\ref{fig2}(a), $\xi_2$ increases by a factor of $6$. At the same time $\xi_1$ changes only by $10\%$. The explanation is that the Fermi velocity of band $2$ is varied in our calculations while $v_{F1}$ is kept constant. For nearly decoupled bands one expects that approximately $\xi_i \propto v_{Fi}$, which was confirmed by the previous calculations within the extended GL approach~\cite{Saraiva2017}. Though this relation is not strictly applicable for finite interband couplings, $\xi_2$ remains more sensitive to changes of $v_{F2}/v_{F1}$ unless $g_{12}$ approaches $g_{11}$~(see below). In Fig.~\ref{fig2}(b) one can see that the increase of $\xi_2$ becomes less pronounced as compared to panel (a) while the increase of $\xi_1$ becomes much more notable: when $v_{F2}/v_{F1}$ varies from $1$ to $5$, $\xi_2$ enlarges by a factor of $3$ whereas $\xi_2$ increases by a factor of $2$. It means that at $g_{12}=0.3g_{11}$ the lengths $\xi_1$ and $\xi_2$ are significantly closer to each other than for the case $g_{12}=0.05g_{11}$. This is further illustrated in Figs.~\ref{fig2}(c) and (d) where the ratio $\xi_2/\xi_1$ is shown versus $v_{F2}/v_{F1}$ for the same two values of the interband coupling. As seen, when $v_{F2}/v_{F1}$ reaches $5$ for the case $g_{12} = 0.05g_{11}$, the ratio $\xi_{2}/\xi_1$ approaches $6$. For $g_{12} = 0.3g_{11}$ we obtain less disparity between the healing lengths, namely, $\xi_{2}/\xi_1 \approx 1.5$ when $v_{F2}/v_{F1}$ reaches $5$.

\begin{figure}[t]
\centering
\includegraphics[width=1\linewidth]{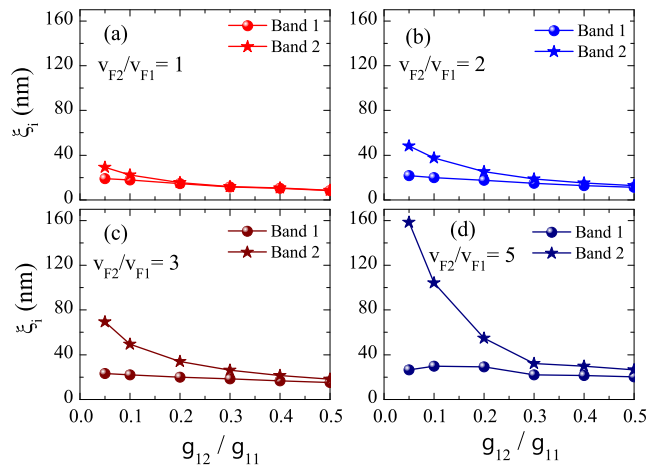}
\caption{(Color online) The condensate healing lengths $\xi_1$ and $\xi_2$ versus the relative interband coupling $g_{12}/g_{11}$ at $T,H=0$ for $v_{F2}/v_{F1}=1$~(a), $v_{F2}/v_{F1}=2$~(b), $v_{F2}/v_{F1}=3$~(c), and $v_{F2}/v_{F1}=5$~(d).}
\label{fig3}
\end{figure}

The dependence of the healing lengths $\xi_1$ and $\xi_2$ on $g_{12}$ is also very sensitive to the value of $v_{F2}/v_{F1}$. This is seen from Fig.~\ref{fig3}, which demonstrates $\xi_i$~($i=1,2$) as functions of the ratio $g_{12}/g_{11}$ for $v_{F2}/v_{F1}=1$~(a), $2$~(b), $3$~(c), and $5$~(d) [solid circles correspond to band $1$ whereas stars are given for band $2$]. One can see in all panels that $\xi_2$ drops significantly with increasing the interband coupling while $\xi_1$ remains almost unaltered. We note that this feature qualitatively agrees with the results of Ref.~\onlinecite{Ichioka2017} obtained by numerically solving the Eilenberger equations.

\begin{figure}[t]
\centering
\includegraphics[width=1\linewidth]{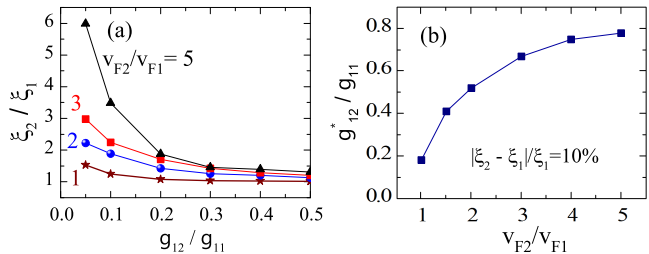}
\caption{(Color online) (a) The ratio $\xi_2/\xi_1$ as a function of $g_{12}/g_{11}$ at $T,H=0$, as calculated for $v_{F2}/v_{F1}=1, 2, 3$ and $5$. (b) The ``length-scales locking" interband coupling $g_{12}^*$ in units of $g_{11}$ versus $v_{F2}/v_{F1}$; the chosen criterion of the locking is taken as $|\xi_2-\xi_1|/\xi_1 \le 0.1$ for $g_{12} > g^\ast_{12}$. }
\label{fig4}
\end{figure}

The dependence of the healing lengths on the interband coupling  is further illustrated in Fig.~\ref{fig4}(a), where the ratio $\xi_2/\xi_1$ is shown versus $g_{12}/g_{11}$ for $v_{F2}/v_{F1}=1,2,3$, and $5$. The difference between $\xi_1$ and $\xi_2$ is more significant for a larger ratio of the band Fermi velocities and for lower values of the Josephson coupling. When $g_{12}/g_{11}$ is sufficiently large, the two healing lengths approach each other, which is known as ``length-scales locking", see Ref.~\onlinecite{Ichioka2017}. This regime reflects the fact that the multiband phenomena are washed out for sufficiently large interband couplings. In this case partial condensates in multiband materials become so strongly coupled that their properties are not distinguished any more. Let us introduce the ``length-scales locking" interband coupling $g_{12}^*$ adopting the criterion $|\xi_2-\xi_1|/\xi_1 \le 0.1$ for $g_{12} > g^\ast_{12}$. [Notice that qualitative conclusions are not sensitive to the particular value in the right-hand side of the inequality for the difference between the two healing lengths.] The dependence of  $g_{12}^*$ on $v_{F2}/v_{F1}$ is illustrated in Fig.~\ref{fig4}(b). One finds that $g_{12}^*$ rapidly increases with $v_{F2}/v_{F1}$ for
$v_{F2} < 4 v_{F1}$ while approaching a saturation for $v_{F2} \gtrsim 5 v_{F1}$. The saturation occurs for $g_{12}^*  \approx 0.8g_{11}$, which is far beyond the regime of nearly decoupled bands.

Based on the results given in Fig.~\ref{fig4}(b), it is also possible to introduce the ``length-scales locking" Fermi velocity ratio $v^*$, below which the difference between $\xi_1$ and $\xi_2$ is negligible. For example, adopting again the locking criterion as $|\xi_2-\xi_1|/\xi_1 \le 0.1$, we find that $v^* \approx 2.0$ for $g_{12} = 0.5g_{11}$ while $v^* \approx 5.0$ for $g_{12} = 0.8g_{11}$.

\subsection{Finite $T$ and zero $H$}\label{sub2}

Let us now discuss how the temperature dependence of the band healing lengths is affected by the ratio $v_{F2}/v_{F1}$. Here the calculations are performed for the same interband couplings as in the previous subsection,  the external magnetic field is zero.

\begin{figure}[t]
\centering
\includegraphics[width=0.9\linewidth]{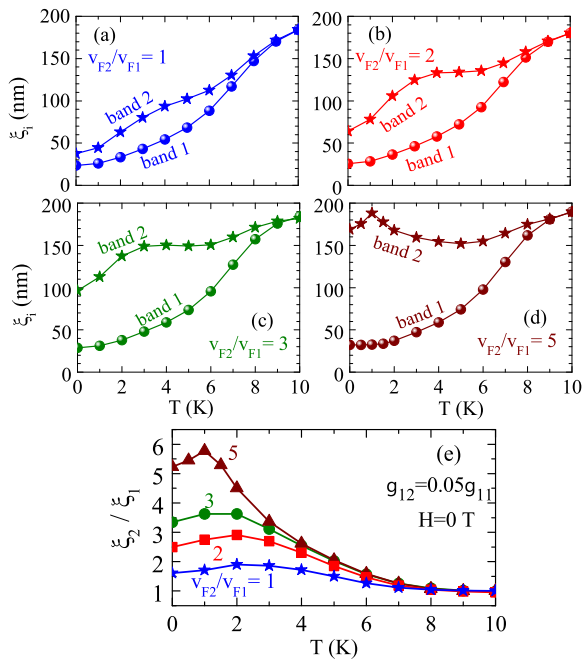}
\caption{(Color online) The healing lengths $\xi_1$ and $\xi_2$  as functions of the temperature ($H=0$) at $g_{12} = 0.05g_{11}$, calculated for  $v_{F2}/v_{F1}=1$~(a), $v_{F2}/v_{F1}=2$~(b), $v_{F2}/v_{F1}=3$~(c), and $v_{F2}/v_{F1}=5$~(d). Panel (e) represents the corresponding ratio $\xi_{2}/\xi_1$.}
\label{fig5}
\end{figure}

In Figs.~\ref{fig5} (a)-(d) one can see the healing lengths $\xi_1$ and $\xi_2$ as functions of the temperature $T$ for $g_{12} = 0.05g_{11}$ and $v_{F2}/v_{F1}=1$, $2$ and $3$ and $5$. As can be expected from our consideration in the previous subsection, $\xi_1$~(circles) exhibits minor variations when passing from (a) to (d) while $\xi_2$ (stars) changes significantly. The reason is mentioned in the discussion of Figs.~\ref{fig2}-\ref{fig4}: $v_{F2}$ is varied in the calculations while $v_{F1}$ is kept constant. The new feature present in the results of Fig.~\ref{fig5} is the nonmonotonic dependence of $\xi_2$ on $T$, clearly seen in the results for $v_{F2}/v_{F1}=2$~(b), $3$~(c), and $5$~(d). This is the effect of the hidden criticality~\cite{Komendova2012} manifesting itself near $T_{c2} = 3.06\,{\rm K}$, where $T_{ci}$ is the critical temperature of the decoupled band $i$. For band $2$, taken as a separate superconductor, the healing length $\xi_2$ increases toward infinity when $T \to T_{c2}$. Though this increase is smoothed and significantly affected by the presence of the interband interactions, its signatures survive at nonzero couplings $g_{12}$. In particular, one observes the plateaus in the temperature dependence of $\xi_2$ in vicinity of $T_{c2}$ in panels (b) and (c). In panel (d) such a plateau disappears in favor of a small but well pronounced peak with the position shifted down to $T=1\,{\rm K}$.

The presence of the hidden criticality is also reflected in the healing lengths ratio $\xi_2/\xi_1$ given versus $T$ in Fig.~\ref{fig5}(e) for the same parameters as in Figs.~\ref{fig5}(a)-(d).  The ratio $\xi_2/\xi_1$ exhibits a maximum for all given values of the Fermi velocities ratio $v_{F2}/v_{F1} = 1$, $2$, $3$ and $5$. The larger is $v_{F2}/v_{F1}$, the higher is the maximal value of $\xi_2/\xi_1$. For example, the maximum $\xi_2/\xi_1$ for $v_{F2}/v_{F1}=5$ is by a factor of $3$ larger than that for $v_{F2}/v_{F1} = 1$. In agreement with the shift down in temperatures of the $\xi_2$-peak in Fig.~\ref{fig5} (d), the peak of $\xi_2/\xi_1$ is also shifted to lower temperatures when increasing $v_{F2}/v_{F1}$. One can also see that $\xi_2/\xi_1$ tends to $1$ as $T$ approaches $T_c$, which is the previously discussed ``length-scales locking" regime near $T_c\approx 10\,{\rm K}$, see Refs.~\onlinecite{Geilikman1967, Koshelev2004, Koshelev2005, Geyer2010, Kogan2011, Shanenko2011, Vagov2012}. For larger values of the ratio $v_{F2}/v_{F1}$, one obtains higher locking temperatures $T^{*}$. The dependence of $T^{*}$ on $v_{F2}/v_{F1}$ and $g_{12}$ is discussed below, at the end of this subsection. Thus, as seen from Fig.~\ref{fig5}, the disparity between $\xi_1$ and $\xi_2$ is the most pronounced for $T \lesssim T_{c2}$ and the maximal value of $\xi_2/\xi_1$ is governed by the hidden criticality.

\begin{figure}[t]
\centering
\includegraphics[width=0.9\linewidth]{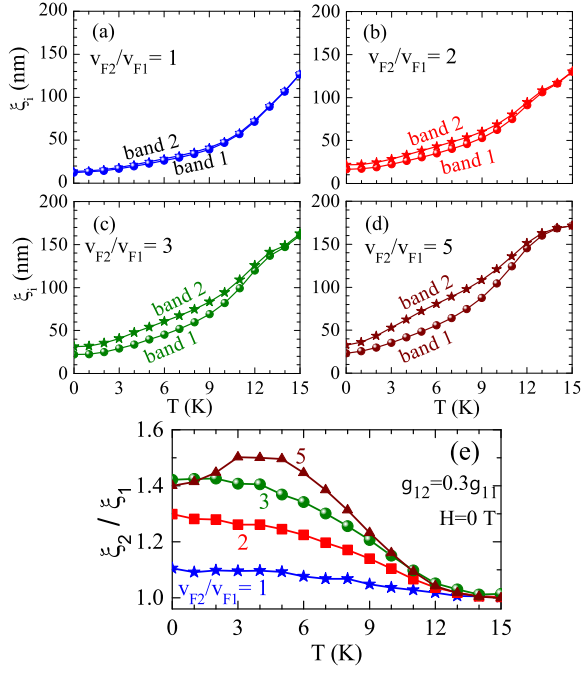}
\caption{(Color online) The same as in Fig.~\ref{fig5},but for the stronger interband coupling $g_{12} = 0.3g_{11}$.}
\label{fig6}
\end{figure}

Now we investigate the healing lengths $\xi_1$ and $\xi_2$ for the significantly larger interband coupling $g_{12} = 0.3g_{11}$. The corresponding temperature dependent results for $\xi_1$, $\xi_2$ and $\xi_2/\xi_1$ are shown in Fig.~\ref{fig6} for the same values of $v_{F2}/v_{F1}$ as in Fig.~\ref{fig5}. One can see that for $v_{F2}/v_{F1}=1$, $2$, and $3$ the healing lengths $\xi_1$ and $\xi_2$ are nearly the same for the whole temperature range $T < T_c$~($T_c \approx 15\,{\rm K}$ for this value of $g_{12}$). For example, when taking the ``length-scales locking" criterion as $|\xi_2-\xi_1| \le 0.1\xi_1$, one finds that for $v_{F2}/v_{F1}=1$, bands $1$ and $2$ are in the locking regime for all temperatures below $T_c$. This agrees with the previous conclusion of Ref.~\onlinecite{Komendova2012} that the effect of the hidden criticality is weakened due to the interband interactions. However, even at the chosen large interband coupling, the signature of the hidden criticality appears again when the band Fermi velocity $v_{F2}$ exceeds $3$-$4\,v_{F1}$. One can see in Fig.~\ref{fig6}(e) that the dependence of $\xi_2/\xi_1$  exhibits a flat maximum, similarly to the case illustrated in Fig.~\ref{fig5}(e). Hence, for the interband coupling $g_{12}=0.3g_{11}$ the maximum in $\xi_2/\xi_1$ is switched on/off by increasing/decreasing the band Fermi velocities ratio. Though the difference between $\xi_1$ and $\xi_2$ is much less pronounced for $g_{12}=0.3g_{11}$ as compared to the results for $g_{12}=0.05g_{11}$, it is far not negligible. In particular, the maximum of $\xi_2/\xi_1$ for $v_{F2}/v_{F1} = 5$ in Fig.~\ref{fig6}(e) is by a factor of $4$ smaller than that in Fig.~\ref{fig5}(e). However, the corresponding difference between $\xi_1$ and $\xi_2$ in Fig.~\ref{fig6}(e) is still notable, being about $50\%$.

\begin{figure}[t]
\centering
\includegraphics[width=1\linewidth]{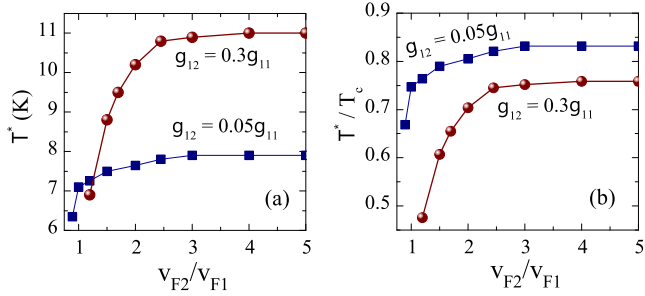}
\caption{(Color online) The length-scales locking temperature $T^*$~(a) and the ratio $T^*/T_c$~(b) as functions of $v_{F2}/v_{F1}$ for the interband couplings $g_{12} = 0.05g_{11}$ and $0.3g_{11}$.}
\label{fig7}
\end{figure}

The last point we address in this subsection, is the effect of the band Fermi velocities on the locking temperature $T^*$. As the locking criterion we again choose $|\xi_2-\xi_1| = 0.1\xi_1$ but now for $T> T^*$. The dependence of $T^*$ on $v_{F2}/ v_{F1}$ is shown in Fig.~\ref{fig7} for the interband couplings $g_{12}= 0.05g_{11}$ and $0.3g_{11}$. In Fig.~\ref{fig7}(a) $T^*$ is given in ${\rm K}$ while the ratio $T^*/T_c$ is demonstrated in Fig.~\ref{fig7}(b). We recall that $T_c$ is not sensitive to the band Fermi velocities and $T_c \approx 10$ K and $\approx 15$ K for $g_{12}=0.05g_{11}$ and $g_{12}=0.3g_{11}$, respectively. As is seen from Fig.~\ref{fig7}, $T^*$ increases with $v_{F2}/v_{F1}$ for either $g_{12} = 0.05g_{11}$ or $0.3g_{11}$. This is due to the fact that the increase of $v_{F2}/v_{F1}$ enlarges the difference between $\xi_1$ and $\xi_2$ at low temperatures, as follows from Figs.~\ref{fig2}-\ref{fig4}. As a result, $\xi_1$ and $\xi_2$ approach each other at larger temperatures, so that  $T^*$ goes closer to $T_c$ when $v_{F2}$ increases. Notice that the intersection of the two curves in Fig.~\ref{fig7}(a) should not lead to any confusion. This does not mean that the locking regime is the same for both interband couplings at the point of the intersection. In particular, this is seen from Fig.~\ref{fig7}(b) where the ratio $T^*/T_c$ is given versus $v_{F2}/v_{F1}$. One can see that $T^*/T_c$ is reduced for $g_{12}=0.3g_{11}$, i.e. the corresponding locking regime is more pronounced, occupying the larger temperature domain in units of $T_c$.

\subsection{Finite $T$ and finite $H$}

\begin{figure}[t]
\centering
\includegraphics[width=0.9\linewidth]{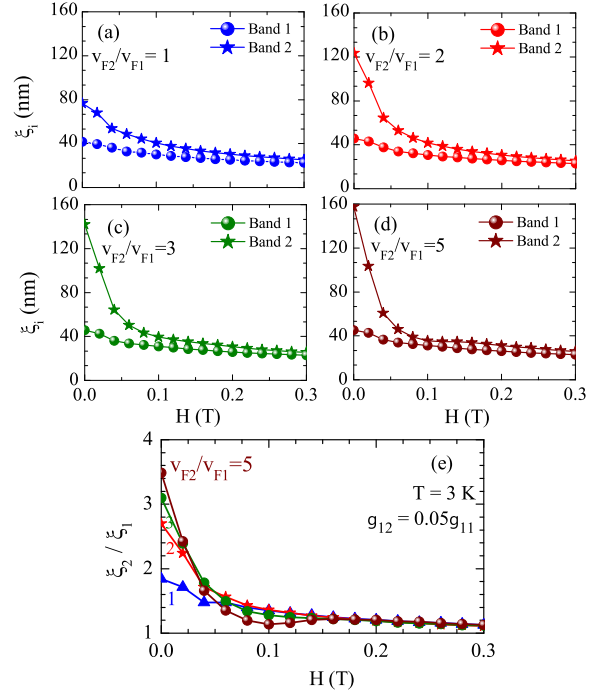}
\caption{(Color online) Healing lengths $\xi_1$ and $\xi_2$ as functions of $H$ for $v_{F2}/v_{F1}=1$~(a),  $v_{F2}/v_{F1}=2$~(b), $v_{F2}/v_{F1}=3$~(c), and $v_{F2}/v_{F1}=5$~(d), calculated at $g_{12} = 0.05g_{11}$ and  $T = 3$ K. Panle (e) shows the corresponding ratio $\xi_{2}/\xi_1$ .}
\label{fig8}
\end{figure}

\begin{figure}[t]
\centering
\includegraphics[width=1.0\linewidth]{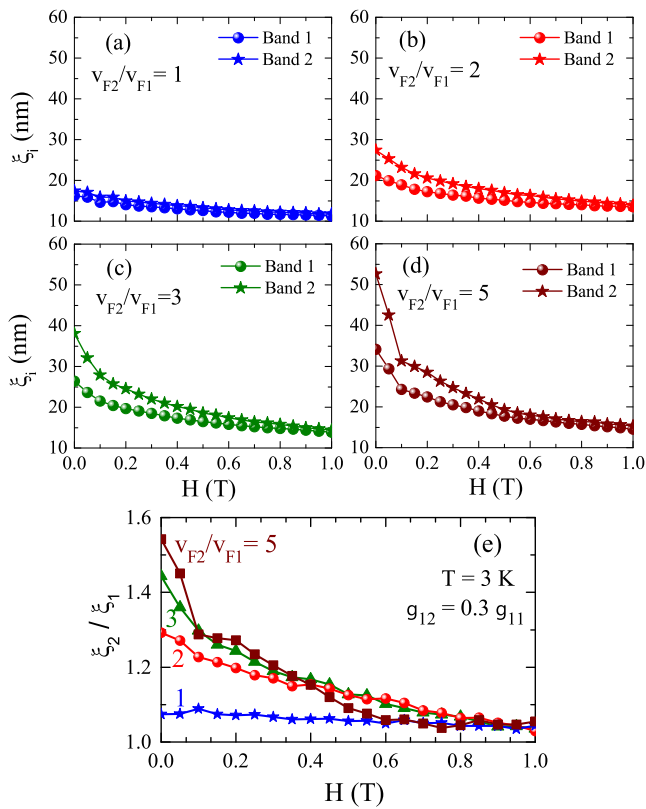}
\caption{(Color online) The same as in Fig.~\ref{fig8} but for the interband coupling
$g_{12} = 0.3g_{11}$.}
\label{fig9}
\end{figure}

Now, let us investigate the healing lengths $\xi_1$ and $\xi_2$ for $H\not=0$. Figures~\ref{fig8}(a)-(d) demonstrate $\xi_1$ and $\xi_2$ as functions of $H$ calculated for the different ratios $v_{F2}/v_{F1} = 1$, $2$, $3$ and $5$ at $g_{12} =0.05g_{11}$ and $T = 3$K. When increasing the external magnetic field, the suppression of the band-dependent gap functions starts near the surface of the cylinder. The region of the suppressed condensate expands and the maximal value of $\Delta_i(\rho)$~(i.e. $\Delta_{i,{\rm bulk}}$) decreases (the condensate is zero at $\rho=0$ and at $\rho=R$). This decrease corresponds to the suppression of the condensate between the densely distributed vortices in the bulk vortex matter near the upper critical field. We recall (see the discussion in the beginning of Sec.~\ref{res}) that the boundary condition $B\to H\not=0$ is suitable to study the healing lengths only in the vicinity of $H_{c2}$.

From Fig.~\ref{fig8}, one can see that the healing lengths are significantly different for $H \to 0$ but this difference disappears when increasing the external field. Both $\xi_1$ and $\xi_2$ monotonically decrease with an increase of $H$ for all values of the band Fermi velocities ratio, which agrees with the results of Ref.~\onlinecite{Ichioka2017}. However, $\xi_1$ is only slightly dependent on $H$ while the decrease of $\xi_2$ is very pronounced. Notice that the isolated vortex also shrinks with increasing the external field, see e.g. Ref.~\onlinecite{Chen2008}.

At high fields, the system approaches the locking regime, which is clearly seen from Fig.~\ref{fig8}(e). When using the locking criterion as $|\xi_2-\xi_1|/\xi_1 \lesssim 0.1$, one obtains  $H^* = 0.27\,H_{c2}$, where $H_{c2} =0.33$ T. The external field at which the vortex solution disappears is interpreted here as the upper critical field. As the boundary condition with a nonzero external field can be relevant only near $H_{c2}$~(see the discussion in the beginning of Sec.~\ref{res}), one can hardly rely upon the obtained value of $H^*$. However, we are able to conclude that near the upper critical field the healing lengths are the same for both contributing condensates notwithstanding the value of $v_{F2}/v_{F1}$.

In Fig.~\ref{fig9} we show $\xi_1$ and $\xi_2$ versus $H$ at the same temperature and values of $v_{F2}/v_{F1}$ as in Fig.~\ref{fig9} but for the interband coupling $g_{12}=0.3g_{11}$. By examining the data in Figs.~\ref{fig9}(a)-(d), we find the same qualitative behavior of the band healing lengths as previously in Fig.~\ref{fig8}. Namely, the band characteristic lengths decrease with increasing $H$ and the disparity between $\xi_1$ and $\xi_2$ becomes more pronounced for larger values of $v_{F2}/v_{F1}$~(at relatively low fields) and less notable for larger $H$. The quantitative results are, of course, different as compared to the case of the weak interband coupling. In particular, by taking the length-locking criterion as $|\xi_2-\xi_1|/\xi_1 \lesssim 0.1$, we find that the band length-scales for $v_{F2}/v_{F1}= 1$ are locked for all magnetic fields. However, taking $v_{F2}/v_{F1} = 2$, $3$ and $5$, we find that the ratio $\xi_2/\xi_1$ becomes smaller than $1.1$ for $H > H^*=0.2\,H_{c2}$, with $H_{c2}=2.7$ T.

Reasonably enough, larger interband couplings shift the locking magnetic field down (as compared to $H_{c2}$). However, we stress again that the boundary condition with a finite external field can be useful only to investigate the healing lengths near $H_{c2}$. In the vicinity of $H_{c2}$ the both healing lengths appear to be the same, irrespective of the particular value of $g_{12}$.

\section{conclusions}
\label{conclusions}

We have studied the effect of the band Fermi velocities on the healing lengths in a two-band superconductor by numerically solving the Bogoliubov-de Gennes equation for a single-vortex solution. Our results demonstrate that near the lower critical field the healing lengths of the two contributing condensates can be significantly different for sufficiently large values of the ratio of the band Fermi velocities $v_{F2}/v_{F1}$. This occurs far beyond the regime of nearly decoupled bands, at the interband couplings up to $g_{12} \sim g_{11},g_{22}$. The most pronounced difference between the healing lengths is observed in the vicinity of or below the hidden critical temperature. The ``length-scales locking" regime takes place near the upper critical field and/or near the critical temperature.

Our study is connected with the long discussion about the possibility to have two coupled condensates with significantly different spatial profiles in the presence of the magnetic effects. Our work clearly demonstrates this possibility for a wide range of the physical parameters. The presence of different healing lengths can significantly change the magnetic response of multiband superconductors as compared to that of single-band ones. For example, it is known that the switching between superconductivity types I and II occurs through the finite intertype domain in the $\kappa$-$T$ plane ($\kappa$ is the Ginzburg-Landau parameter), see e.g. Refs.~\onlinecite{Brandt2011} and \onlinecite{Vagov2020}. It has been proved~\cite{Vagov2016,Cavalcanti2020} that this domain significantly enlarges in two-band and multiband superconductors (with respect to the single-band case) if the healing lengths of different contributing condensates are significantly different. Our present study is a solid compliment to these previous investigations based on the perturbation theory in the vicinity of $T_c$. We confirm that multiband materials with significantly different band Fermi velocities are most promising in searching for unconventional superconducting magnetic properties because of the presence of multiple condensates governed by different spatial scales.

\acknowledgements

This work was supported by Natural Science Foundation of Zhejiang Province (Grant No. LY18A040002), Science Foundation of Zhejiang Sci-Tech University(ZSTU) (Grant No. 19062463-Y) and National Natural Science Foundation of China (Grant No. NSFC-11375079). Y. C. and H. Z. acknowledge the hospitality of the Physics Department of the Federal University of Pernambuco during their visit.

\bibliographystyle{apsrev}

\end{document}